\documentclass[12pt]{article}
\input epsf
\newcommand{\be}{\begin{equation}}
\newcommand{\ee}{\end{equation}}
\newcommand{\bea}{\begin{eqnarray}}
\newcommand{\eea}{\end{eqnarray}}
\newcommand{\ba}{\begin{array}}

\newcommand{\ea}{\end{array}}

\def\bbox{{\,
\lower0.9pt\vbox{\hrule \hbox{\vrule height 0.2 cm
\hskip 0.2 cm \vrule height 0.2 cm}\hrule}\,}}
\newcommand{\dsl}{\pa \kern-0.5em /}

\newcommand{\nn}{\nonumber \\}

\def\ds{\raise.15ex\hbox{/}\kern-.57em\partial}
\def\Ds{\,\raise.15ex\hbox{/}\mkern-13.5mu D}
\begin{document}

\baselineskip 18pt


\begin{titlepage}
\vfill
\begin{flushright}
SNUTP01-030\\
KIAS-P01040\\
hep-th/0109217\\
\end{flushright}

\vfill

\begin{center}
\baselineskip=16pt
{\Large\bf DLCQ of Fivebranes, Large $N$ Screening,
and \\$L^2$ Harmonic Forms on Calabi Manifolds}
\vskip 10mm
{Chanju Kim,$^{1}$ Kimyeong Lee,$^{2}$ and Piljin Yi $^{3}$}
\vskip 8mm
{\small\it
$^1$ School of Physics and Center for Theoretical Physics\\
Seoul National University, Seoul, 151-747,
Korea \\
\vskip 3mm
$^{2,3}$ School of Physics, Korea Institute for Advanced Study\\
207-43, Cheongryangri-Dong, Dongdaemun-Gu, Seoul 130-012, Korea}
\end{center}
\vskip 10 mm
\par
\begin{center}
{\bf ABSTRACT}
\end{center}
\begin{quote}
We find one explicit $L^2$ harmonic form for every Calabi manifold.
Calabi manifolds are known to arise in low energy dynamics of solitons
in Yang-Mills theories, and the $L^2$ harmonic form corresponds
to the supersymmetric ground state.  As the normalizable ground state
of a single $U(N)$ instanton, it is related to the bound state of a
single D0 to multiple coincident D4's in the non-commutative setting, or
equivalently a unit Kaluza-Klein mode in DLCQ of fivebrane worldvolume
theory.  As the  ground state of  nonabelian massless monopoles
realized around a monopole-``anti''-monopole pair, it shows how the
long range force between the pair is screened in a manner reminiscent
of large $N$ behavior of quark-anti-quark potential found in AdS/CFT
correspondence.

\end{quote}

\vfill
\vskip 5mm
\hrule width 5.cm
\vskip 5mm
\begin{quote}
{\small
\noindent
$^1$ E-mail: cjkim@phya.snu.ac.kr\\
$^2$ E-mail: klee@kias.re.kr\\
$^3$ E-mail: piljin@kias.re.kr\\
}
\end{quote}
\end{titlepage}
\setcounter{equation}{0}

\section{Introduction}

Several recent research issues have come together to the study of  the
supersymmetric quantum dynamics on a family of hyper-K\"ahler manifolds,
one for each $4n+4$ space with nonnegative integer $n$.
These manifolds, the so-called Calabi manifolds \cite{calabi}, arise naturally
as the moduli spaces of  certain configurations of magnetic monopoles
\cite{LWY} or a single instanton  on a noncommutative space \cite{LY}.

The Calabi manifold is perhaps  the simplest class of nontrivial
hyper-K\"ahler manifolds. The Calabi manifold of dimension $4n+4$ is
$T^*CP^{n+1}$, the cotangent bundles of the complex projective space
$CP^{n+1}$, equipped with a hyper-K\"ahler metric of cohomogeneity one
and  the triholomorphic isometry $SU(n+2)$ \cite{DS}.  There is one
scaling coordinate for the size of the orbit and the generic orbit of
the isometry is the coset space $SU(n+2)/U(n)$ of dimension $4n+3$,
which collapses to a $CP^{n+1}$ bolt in the minimal scaling.

One form of the metric is
\begin{equation}
C_{AB}d{\vec x}_A\cdot d{\vec x}_B + (C^{-1})_{AB} (d\psi_A+\vec \omega_{AC}
\cdot d\vec x_C)(d\psi_B+\vec \omega_{BD} \cdot d\vec x_D). \label{calabi}
\end{equation}
The coordinate $\psi_A$ has the period $4\pi$ and the symmetric matrix
$C_{AB}$ has the form,
\begin{equation}
C_{AB}=\left(\frac{\delta_{AB}}{|\vec x_A|} + \frac{1}{
|\sum_{E} \vec x_E - \vec R\;|}\right) .
\end{equation}
while the vector potentials $\vec \omega_{AB}$ solves the equations
\begin{equation}
\vec \nabla_D \,C_{AB}=\vec \nabla_D \times \vec \omega_{AB} .
\end{equation}
This form of metric naturally arises as a hyper-K\"ahler reduction of
flat $R^{4n+8}$ \cite{GRG}. The hyper-K\"ahler reduction preserves some of the
isometries: More specifically, there is an $SU(n+2)$ isometry which
preserves not only the metric but also all three k\"ahler forms, while
an additional $U(1)$ rotating $\vec x_A$'s orthogonally to $\vec R$,
preserves the metric and one k\"ahler form. Such isometries are called
triholomorphic and holomorphic, respectively.

Recently, there has been an interesting progress in the study of
the Calabi manifold. Ref.~\cite{CGLP} presented an alternate form 
of the metric, which was then used to find the explicit
expression of an $L^2$ harmonic form for eight and twelve dimensional 
manifolds. Here, we use this new form for the metric to find the
explicit expression of one normalizable harmonic form on the
Calabi manifold of arbitrary dimensions. 
The $L^2$ harmonic form is anticipated to be unique on each Calabi
manifold \cite{Hitchin}, and therefore must be self-dual or anti-self-dual
middle form. We indeed find such an
$L^2$ harmonic form. 

We will then explore the physical implications in two physical settings;
in the context of (2,0) theories in six dimensions 
\cite{witten,seiberg,andy6d}  and  in the context of the large $n$
dynamics of monopole and anti-monopole in partially broken gauge theories.
Recall that a normalizable harmonic form, if it exists, corresponds to a 
normalizable ground
state wave function of the Hamiltonian with eight supersymmetries.  
This middle form has several physical meanings.  The four
dimensional case with $n=0$ is the so-called Eguchi-Hanson metric
\cite{EH}. The normalizable middle form in this space has been found
and studied in many places \cite{HIT,LY,TONG}.

Initially, Calabi discovered this family of metrics as a study of
the hyper-K\"ahler space. It has also appeared in the study of the
moduli space of magnetic monopoles and instantons.
The simplest example is the moduli space of a single instanton in
$U(n+2)$ theory \cite{LY}. Its moduli space in the center of mass frame is
$4(n+1)$ dimensional hyper-K\"ahler space with one scale parameter and
the principal orbit $SU(n+2)/U(n)$. This space is singular at the zero
scale size, which can be blown up to a finite bolt, making the space
nonsingular. This has been done in the ADHM formalism. This blow-up of
the instanton moduli space can be achieved physically by going to the
non-commutative space, which is equivalent to introducing a FI term on
the ADHM formalism.

Recall that $U(m)$ type (2,0) theories arise as low energy dynamics
of $m$ coincident fivebranes~\cite{andy6d}.
Alternatively, it also
arises as the strong coupling limit of the five dimensional $U(m)$
Yang-Mills theories with 16 supersymmetries. Not much is known about
these theories beyond their existence and somewhat tentative
statements about their anomaly structure \cite{mans,anomaly,Intriligator}.
The only explicit approach
to these theories, known to date, is the DLCQ limit thereof \cite{ofer}.
Because
instanton solitons of five-dimensional Yang-Mills theory carries the
Kaluza-Klein momentum modes along the extra circle, DLCQ of the (2,0)
theories have partons that are really instanton solitons. This
approach to the (2,0) theories is thus naturally tied to instanton
dynamics of five dimensional Yang-Mills theories. An early study of
the relationship was conducted by Aharony et.al. \cite{ABS},
who tried to extract
spectrum of chiral primaries of (2,0) theories from cohomology
counting on the instanton moduli space. While they found numerous
nontrivial cohomology generators, only one of them is associated with
free center of mass degrees of freedom and may
correspond to $L^2$ harmonic form on Calabi space. Our findings here show
that indeed one such $L^2$ harmonic form exists for any number of
coincident fivebranes. Extrapolation of this correspondence to
arbitrary number $k$ of instantons and arbitrary $m$, gives us a
conjecture that there should be a unique $L^2$ harmonic form for each
and every pair $m$ and $k$. Our finding in this paper supports the
conjecture by explicitly constructing such harmonic forms for all $n$
and $k=1$.\footnote{In the paper~\cite{TONG} by one of us, the moduli
space of two $U(1)$ instantons is shown to be Eguchi-Hanson which
possesses as a unique normalizable middle form in its moduli
space. This state was interpreted as the mode for two momenta, giving
an independent check for the $k=2,m=1$ case.}

Calabi manifold and the $L^2$ harmonic form on it find another application
in the moduli space dynamics of certain non-Abelian monopoles \cite{solution}.
Consider $N=4$ $SU(n+4)$ theory broken to $U(1)^{n+3}$, which contains
$n+3$ distinct types of fundamental monopoles.  The moduli space of these
$n+3$ distinct monopoles is explicitly known \cite{many}.
The moduli space dimension is $4(n+3)$ and each four degrees correspond to
the position and phase of each fundamental monopole.  When the gauge
symmetry is partially restored to $U(1)\times SU(n+2)\times U(1)$ so
that only two of them at the end of the Dynkin diagram remain massive,
the net magnetic charge is orthogonal to unbroken $SU(n+2)$
generators. In this limit the moduli space remains still sensible 
and still is of $4(n+3)$ dimensions \cite{LWY}. 

This moduli space turns out to have two scale parameters, one for the 
relative distance between two massive monopoles and one for the scale 
parameter for the massless monopole clouds surrounding two massive monopoles. 
This latter corresponds to the sum of the relative distances between pairs of
interacting monopoles, some of which is now massless can no longer be regarded
as isolated solitons. In such a massless limit, only the two scales are 
invariant under global gauge rotations and spatial rotations. All the rest
are associated with $U(1)\times SU(n+2)\times U(1)$ 
or $SO(3)$ rotation of the solution. 
The relative moduli space of dimension $4(n+2)$
has $U(n+2)$ triholomorphic isometry and $SU(2)$ rotational symmetry
which is just holomorphic with respect to one of three complex structure.
The particle dynamics on such a moduli space has been studied extensively 
in a recent paper \cite{chen}. 

When we take a further limit of two massive particles becoming
infinitely massive so that
their relative distance is fixed, or after the hyper-K\"ahler quotient
of the Taubian-Calabi metric with the $U(1)$ triholomorphic isometry,
the resulting $4(n+1)$ dimensional metric is the Calabi metric which
we study.
Since magnetic monopoles in the $N=4$ supersymmetric four
dimensional Yang-Mills Higgs theories have half supersymmetry, the
moduli space metric has the eight supersymmetries.

One of the key achievement in the AdS-CFT correspondence between the
classical supergravity theory on the $AdS_5\times S^5$ and $N=4$
supersymmetric Yang-Mills theory is the calculation of the potentials
between quarks in the large $e^2 N$ limit, keeping $e^2$ small
\cite{RY,Maldacena}.  The non-analytical expression $\sqrt{e^2 N}$ has
been found as the coefficient. There has been several works in the
field theory to reproduce this highly nontrivial
result~\cite{GORDON}. The present  work is partially motivated by this
result. Here we study the potential in the large $N$ limit. As the
non-perturbative reduction of the attractive quarks is happening in the
low energy when the overall coefficient is quite small, one may expect
that the low energy dynamics we study here may capture the right
physics. As we will see our result is encouraging but not identical
to the AdS-CFT calculation. One may regard our work as a first attempt
to understand the AdS-CFT result in our direction.

In Section 2, we consider DLCQ of fivebranes and show how Calabi manifold 
appears naturally in that context. By turning on non-commutativity, we
may reduce the quantum mechanics to quantum mechanics on 
the resolved instanton moduli space of 
$U(m)$ theory. The latter is nothing but the Calabi manifold.
In Section 3, we study the metric of the Calabi space introduced
recently.  There we find the explicit expression for the normalizable
middle forms for each $n$.
In Section 4, we interpret the harmonic form as corresponding to unit
Kaluza-Klein mode associated with free center of mass motion of coincident
fivebranes. 
In Section 5, we study an unrelated application of the harmonic form.
We consider static potential between a pair of non-Abelian monopoles which
can be regarded as a monopole-anti-monopole pair.
We find an screening effect at large $n$,
reminiscent of a similar effect observed in AdS/CFT setting, even though 
we are working at the level of low energy dynamics only.
In Section 6, we conclude with a summary.

\section{DLCQ of Fivebrane Theory and Instanton}

Instanton of five dimensional Yang-Mills theories plays the role of
partons when one considers DLCQ limit of coincident fivebranes.
For each instanton number, one is probing dynamics with fixed total
Kaluza-Klein momentum sector, and this correspondence predicts
existence of certain bound state on DLCQ description. In particular,
given any number of fivebranes, there is always a free center of mass part of
the world-volume dynamics, given by a single tensor multiplet. These
degrees of freedom must manifest themselves as normalizable supersymmetric
ground states. As we explain below, the $L^2$ harmonic form we
found provide exactly such state, and gives us a consistency check
on the DLCQ prescription.

The actual DLCQ quantum mechanics is some Yang-Mills type quantum
mechanics with 8 supercharges, Higgs branch of which corresponds to
instanton moduli space while Coulomb branch corresponds to freed D0
branes in spacetime \cite{ofer}. Because we are dealing with quantum mechanics,
a clean separation between branches is not possible, and in general
one must consider the full quantum mechanics in order to understand
dynamics of fivebrane world-volume \cite{d1d5,micha}.
On the other hand, problem gets quite
simplified when non-commutativity is turned on. This removes the
Coulomb branch altogether, and leaves behind a resolved version
of the instanton moduli space as the Higgs branch.\footnote{
Refs.~\cite{H} addressed the question of whether a single D0 binds
to a single D4 without any such deformation. Generalization of this 
approach to arbitrary number of D4 is significantly more difficult
and remains an open problem.} With this,
the system may be reduced to sigma model onto the resolved instanton
moduli space, which we will presently show to be exactly the Calabi
manifold. Here we will use the result of Ref.\cite{LY}, and
present the moduli space of non-commutative
 instanton defined on $R^3\times S^1$
and then derive Calabi manifold as instanton moduli space on
$R^4$ by taking the decompactification limit..

Consider a periodic instanton on $R^3\times S^1$, of non-commutative
$U(n+2)$ theory in $4+1$ dimensions. Such a periodic instanton is known
to consist of $n+2$ partons which are BPS monopoles \cite{caloron}. Using this
realization, the instanton moduli space has been computed. The monopoles
involved are all distinct, and the metric of low energy dynamics
is toric with $n+2$ $U(1)$ triholomorphic isometries. Only interaction
between distinct monopoles is due to exchange of purely $U(1)$ massless
vector multiplets, which comes in a scale invariant form. This
uniquely fixes the moduli space metric.

Separating out
flat $R^3\times S^1$ that corresponds to pure translational degrees of
freedom, the moduli space is a $4n+4$ dimensional hyper-K\"ahler space.
Let $l$ be radius of $S^1$ and $e$ the five dimensional Yang-Mills
coupling. The metric can be written as
\begin{equation}
{\cal G}=\frac{4\pi^2 l}{e^2}\,\left(
C_{AB}d\vec x_A\cdot d\vec x_B + (C^{-1})_{AB} (d\psi_A+\vec \omega_{AC}
\cdot d\vec x_C)(d\psi_B+\vec \omega_{BD} \cdot d\vec x_D)\right) .
\end{equation}
$\psi_A$ are periodic in $4\pi$, and the symmetric matrix $C_{AB}$ has the
form,
\begin{equation}
C_{AB}=\left(\mu_{AB} + \frac{\delta_{AB}}{|\vec x_A|} + \frac{1}{
|\sum_{E=1}^{n+1} \vec x_E - 2\pi\vec\zeta/l\;|}\right) .
\end{equation}
The hyper-K\"ahler property requires
the vector potentials $\vec \omega_{AB}$ to be related to $C_{AB}$
by
\begin{equation}
\vec \nabla_D \,C_{AB}=\vec \nabla_D \times \vec \omega_{AB} .
\end{equation}
The constant matrix $\mu$ is proportional to
the reduced mass matrix of the partons, and is
determined uniquely by the Wilson line along $S^1$. In particular, it
is non-degenerate in the maximally broken phase, and vanishes identically
when the gauge symmetry is completely restored. Three vectors $\vec x$
are relative positions of the partons in $R^3$.

Non-commutativity is encoded in the vector $\vec\zeta$, which
is the anti-self-dual part of the
commutator, $\theta_{\mu\nu} =i[x_\mu,x_\nu]$, which defines the
non-commutativity of the $R^3\times S^1$. The anti-self-dual part
is effectively a vector under $SO(3)$ of $R^3$.

Let us take the limit where the $U(n+2)$ gauge symmetry is restored by
turning off Wilson line. In effect, we set $\mu=0$.
\begin{equation}
C_{AB}= \left(\frac{\delta_{AB}}{|\vec x_A|} + \frac{1}{
|\sum_{E=1}^{n+1} \vec x_E - 2\pi\vec\zeta/l\;|}\right).
\end{equation}
Note that, apart from an overall multiplicative scale, 
corresponding metric is exactly that of Eq.~(\ref{calabi}). Thus
the instanton moduli space in the limit of vanishing Wilson line
is a Calabi manifold.

There is a further limit we can take while maintaining Calabi manifold
as the moduli space. We may choose to go back to $R^4$ by sending $l$
to infinity. This may seem like a singular limit, given that $\zeta/l$
would vanish. However, such a limit should describe a
resolved instanton moduli space which is smooth. The resolution of the
apparent conflict lies in the fact
that  there is an overall $l$ multiplying the metric,
so in order to reach a finite limit, one must rescale the collective
coordinates by $\vec x\rightarrow \vec x/l$, upon which the
moduli space remains Calabi manifold with $\vec R=2\pi\vec\zeta$.

Thus, an instanton on non-commutative $R^3\times S^1$ or on non-commutative
$R^4$ is characterized by the common, nontrivial part of the moduli space
given by the Calabi manifold. The only difference lies in the flat,
center of mass part of the moduli space, $R^3\times S^1$ and $R^4$,
respectively. In next section, we will introduce a new coordinate system,
where quantization procedure proves a bit easier, and find a
normalizable, bound state at zero excitation energy. {\it
Thus we effectively will have shown the existence of the
Kaluza-Klein mode associated with the free part of the fivebrane
theory for any number of fivebranes.} Unlike the case of the matrix theory
in the bulk \cite{bfss,d0}, this is a rather nontrivial test even for the unit
Kaluza-Klein mode, since the parton itself carries many internal
degrees of freedom which must be quantized.

\section{$L^2$ Harmonic Form on Calabi Manifolds}

The above metric, although quite simple, is somewhat unwieldy for two reasons:
neither the complex structures nor the $SU(n+2)$ isometry is easy to
see in this form. An alternate form was recently found, which take
advantage of the principal orbits \cite{CGLP}. In this second version, the
metric is written in terms of a single radial coordinate $\rho$ and
a 1-form frame on $S^{4n+3}$, $\sigma_a$, $\bar\sigma_a$, $\Sigma_a$,
$\bar\Sigma_a$, $\nu$,  $\bar\nu$, and $\lambda$;
\be
h(\rho)^2\,d\rho^2+a(\rho)^2\sum_{a=1}^{n}
|\sigma_a|^2 +b(\rho)^2\sum_{a=1}^{n}
|\Sigma_a|^2+c(\rho)^2|\nu|^2+f(\rho)^2\lambda^2,
\ee
with appropriate functions
\be
a(\rho)^2=(\rho^2-1)/2,\;
b(\rho)^2=(\rho^2+1)/2,\;
c(\rho)^2=\rho^2,\;
\ee
and
\be
h(\rho)^2=\rho^4/(\rho^4-1), \;
f(\rho)^2=\rho^2(1-1/\rho^4)/4.
\ee
This metric is equivalent to the above up to an overall rescaling of
the metric, upon relating the radial coordinate $\rho$ with
$\vec x_A$ by
\be
\rho^2=\left((\sum_A |\vec x_A|)+|\sum_A \vec x_A- \vec R\,|\right)/R.
\ee

\subsection{Invariant Form on Calabi Manifold}

The manifold
comes with a triholomorphic $SU(n+2)$ isometry, and also a $U(1)_\lambda$
isometry whose dual is $f^2\lambda$. Under the $U(1)_\lambda$ isometry,
whose Killing  vector is dual to $f^2\lambda$, the complex 1-forms are
charged as,
\bea
\sigma_a &\rightarrow & 1/2,\nn
\Sigma_a &\rightarrow & -1/2,\nn
\nu &\rightarrow & 1.
\eea
Alternatively, we may view $S^{4n+3}$ as a deformed version of
coset $SU(n+2)/U(n)$ where $U(n)$ acts on the right. In this latter
viewpoint, the above 1-forms are part of left-invariant 1-form basis
of $SU(n+2)$ group manifold, transforming as adjoint under the right
action of $SU(n+2)$. In particular, the above 1-forms are grouped
into
\bea
\sigma_a &\rightarrow & [n]_{1},\nn
\Sigma_a &\rightarrow & [n]_{1},\nn
\nu &\rightarrow&  [1]_0,
\eea
under the $U(n)$ as the subgroup of right $SU(n+2)$.

For convenience, we will define the complex orthonormal basis
\bea
s &\equiv& h\,d\rho+if\,\lambda, \nn
p &\equiv& c\,\nu, \nn
\xi_k &\equiv& a\,\sigma_k, \nn
\zeta_k &\equiv& b\,\Sigma_k,
\eea
with the corresponding conjugated ones denoted by an overbar.

The manifold is hyper-K\"ahler, which means that there are three covariantly
constant K\"ahler forms,
\bea
K^{(1)}&=&s\bar s +p\bar p +\sum_k \xi_k\bar \xi_k -
\sum_k\zeta_k\bar\zeta_k, \nn
K^{(2)}+ iK^{(3)} &=&sp+i\sum_k \xi_k\bar\zeta_k.
\eea
The associated complex structures $J^{(a)}$ form an $SU(2)_J$ algebra, which
is an R-symmetry of the supersymmetric quantum mechanics onto the Calabi
manifold. Under this $ SU(2)_J$, the 1-forms are split into $(2n+2)$ doublets,
which are
\be
\left(\begin{array}{c}\xi_k \\ i\zeta_k \end{array}\right),
\ee
and
\be
\left(\begin{array}{c}i\bar \zeta_k\\ \bar \xi_k \end{array}\right),
\ee
for each $k=1,...,n$, and also
\be
\left(\begin{array}{c}s\\ \bar p \end{array}\right),
\ee
and
\be
\left(\begin{array}{c}-p \\   \bar s \end{array}\right).
\ee

We will be looking for a normalizable wavefunction of the complex
$N=4$ supersymmetric quantum mechanics. In other words, we will search
for $L^2$ harmonic forms on the Calabi manifold. Such a harmonic form
is anticipated to be unique on Calabi manifold \cite{Hitchin},
and we will assume that this is the case throughout the paper.

The uniqueness imposes several constraint immediately.
First, it should be either self-dual or anti-self-dual form, which also
implies that it is a middle-dimensional form. Second, the uniqueness implies
that it is a singlet under $U(n)$ and  $U(1)_\lambda$ and also under
the $SU(2)_J$ R-symmetry. The latter, in particular, proves to be very
constraining. It implies that the middle form is of the form,
\be \label{middle}
\Psi=\sum F_a(\rho)\Omega_a,
\ee
where middle forms $\Omega_a$'s are constructed out of the orthonormal
basis above and invariant under $U(n)\times U(1)_\lambda\times SU(2)_J$.

After playing with the basis, one can see that any such
invariant middle $\Omega_a$ can be constructed by multiplying the following
2-forms and 4-forms,
\bea
A &\equiv& s\bar s - p\bar p, \nn
B &\equiv& \sum_k \xi_k \bar\xi_k+ \zeta_k \bar\zeta_k, \nn
C & \equiv &-2isp\sum_k \zeta_k\bar\xi_k -2i \bar s \bar p
\sum_k \xi_k\bar\zeta_k +
(s\bar s+ p\bar p)\sum_k(\xi_k \bar\xi_k - \zeta_k \bar\zeta_k), \nn
D &\equiv & -\sum_k \xi_k\bar\zeta_k \sum_k \zeta_k\bar\xi_k
+\sum_k \xi_k\bar\xi_k \sum_k \zeta_k\bar\zeta_k .
\eea
It should be immediately clear that they are invariant under $U(n)$
and also under $U(1)_\lambda$. Invariance under $SU(2)_J$ needs a little bit
more of scrutiny. $A$ and $B$ are constructed using the usual invariant
bilinear
form of $SU(2)_J$ while $C$ and $D$ are constructed by first forming
a pair of $SU(2)_J$ triplets that are also invariant under $SU(n)$, and
then multiplying two such to form a singlet. We could have used the totally
antisymmetric tensor of $SU(n)$ to construct invariant forms, but
when we require $SU(2)_J$ invariance of the final middle form, they
can be rewritten in terms of the above four. In $4n+4$ dimensions, 
invariant middle forms are,
\bea
&& D^l B^{n-2l+1},\nn
&& D^l A B^{n-2l},\nn
&& D^l A^2 B^{n-2l-1},\nn
&& D^l C B^{2-2l-1},
\eea
with $l=0,1,2,\dots$ etc. There is
a crucial identity among the first set that goes as
\be \label{identity}
0=\sum_{l=0}^{[n/2]} (-1)^l \left(\begin{array}{c}n-l+1  \\ l\end{array}\right)
D^l B^{n-2l+1}.
\ee
which comes about due to the antisymmetric nature of the wedge product and 
the fact that we are forming singlets by multiplying $n$ doublets $n+1$ 
times.\footnote{See Appendix B for more detail.}

\subsection{$L^2$ Harmonic Form}

In principle, the self-dual middle form $\Psi$ is harmonic if and only if it
is closed,
\be
d\Psi=0.
\ee
On the other hand, invariance under $SU(2)_J$ implies that the middle form is
of type $(n+1,n+1)$ with respect to any Hodge decomposition. This 
 implies that $\Psi $ is closed if and only if is is closed
under holomorphic exterior derivative,
\be
\partial \Psi = 0.
\ee
We will solve this equation.
For this, we need the exterior algebra for the left-invariant 1-forms on
$SU(n+2)$ which is summarized in Appendix A.
Since the middle form $\Psi$ is entirely constructed in terms of the forms
$A,\ B,\ C$ and $D$, it is convenient to have their derivatives expressed
in terms of them. From the exterior algebra given in Appendix A, we find
\bea \label{exterior}
\partial A &=& \frac{4f}{c^2} sA - \frac{1}{2fc^2} sB - \frac{1}{2f}E, \nn
\partial B &=& \frac{1}{2f} sB + \frac{1}{2fc^2}E, \nn
\partial C &=& -\frac{1}{2f} sB^2 - \frac{3}{2fc^2} sAB + \frac{2}{f}sD
               -\frac{3}{2f}EA - \frac{1}{2fc^2}EB, \nn
\partial D &=& \frac{1}{f} sD ,
\eea
where we introduced a 3-form
\be
E = 2i \bar p \sum_k \xi_k \bar\zeta_k
          + s \sum_k (\xi_k \bar\xi_k - \zeta_k \bar\zeta_k),
\ee
which has spin $+1/2$ under $SU(2)_J$. Then, $\partial \Psi$ is written
as a polynomial of $A$, $B$, $C$, $D$ and $E$, and the harmonicity condition
$\partial \Psi = 0$ means that the coefficients of independent monomials
should all vanish. The resulting equations form a set of first
order differential equations for the coefficient functions $F_a(\rho)$ in
(\ref{middle}) and many additional algebraic relations. In addition, 
from the self-duality condition
of $\Psi$ we will have further algebraic relations.
Solving directly all these differential and algebraic equations, in
principle, one can obtain the middle form, though it would not be a
simple task in general $4n+4$ dimensions.

Given the uniqueness of the solution, however, the actual route we
take is to work with a plausible ansatz which is chosen after some
experience with a few low dimensional cases up to, say, 20 dimensions ($n=4$).
Remarkably, it turns out that the coefficient functions of the middle form
are not independent but may be expressed in terms of two functions
in the following way,
\be
\Psi = F_n(\rho) \sum_l (a_l A^2 + b_l B^2 + c_l C) D^l B^{n-2l-1}
      + G_n(\rho) \sum_l d_l A D^l B^{n-2l},
\ee
where
\bea
F_n(\rho) &=& \frac{1}{\rho^2(\rho^2+1)^{n+2}}, \nn
G_n(\rho) &=& \frac{1}{\rho^4(\rho^2+1)^n} + nF_n(\rho),
\eea
and $a_l,\ b_l, c_l$ and $d_l$'s are numerical coefficients to be determined
from the conditions of harmonicity and self-duality for $\Psi$. Here,
$b_0$ may be put to be zero using the identity (\ref{identity}).
Inserting this ansatz into $\partial \Psi = 0$ and using (\ref{exterior}),
we obtain many differential
and algebraic consistency equations for $F_n(\rho),\ G_n(\rho)$ and other
coefficients.
Eventually, with the above form of $F_n(\rho)$ and $G_n(\rho)$, all
the differential equations reduce to algebraic ones which have a unique
solution up to an overall normalization,
\bea \label{coefficient}
a_l &=& (n-2l)(n-2l+1)U_l,\nn
b_l &=& -2l U_l,\nn
c_l &=& a_l, \nn
d_l &=& (n-2l+1)U_l ,
\eea
where
\be
U_l = (-1)^l \pmatrix{n-l+1 \cr l}.
\ee
Note that these coefficients are  similar to those in
Eq.~(\ref{identity}) which is crucially used in the calculation.
In addition, it can be shown that the resulting $\Psi$ is indeed self-dual,
the proof of which is given in Appendix B. This solution is normalizable
as we see from the form of $F_n(\rho)$ and $G_n(\rho)$. Also, for $n \le 2$,
it reduces to the corresponding $L^2$ harmonic form found in \cite{CGLP}.

\section{The Ground State of Instanton and Free Motion of Fivebranes}

The $L^2$ harmonic form of previous section finds an obvious
interpretation here in the context of DLCQ of fivebrane theory.
One  set of states on fivebranes
that separates out from the rest is the free center of mass
degrees of freedom which consist of 5 scalars, 4 symplectic-majorana
spinors, and one chiral tensor multiplet. This multiplet is
composed of 16 degrees of freedom. One the other hand, an
$L^2$ ground state of instanton comes with additional degeneracy
due to the superpartner of free, so far neglected, $R^4$ part of moduli
dynamics. There are 4 free complex fermions, which induces degeneracy
of 16, exactly the right amount to form the tensor multiplet.\footnote{
An interesting question is how other states
from interacting part of fivebrane theory, namely
interacting (2,0) theory. Aharony et.al \cite{ABS} argued that the relevant
quantity is the compact  cohomology of the total instanton moduli space.
Translated to compact cohomology of the relative part of the instanton
moduli space, the statement becomes
\be
H^{2n+2l}_{compact}=Z,
\ee
for $l=1,\dots,n+2$. With the single exception of $l=1$ case
which we already interpreted as  coming from free part of (2,0) theory,
it is not clear if any of these
cohomology generators can be represented by an harmonic form.}

A consistency check of this interpretation can be found when
we separate fivebranes from each other. The counting of the
normalizable ground state in fact changes dramatically in this case.
This induces a potential to the moduli space dynamics and the
counting of the ground state becomes a good deal easier. For $m=n+2$
separated fivebranes, the number of $L^2$ ground states is
precisely $m$ \cite{LY,index}.
On the other hand, in this phase, each and every fivebrane is
described by a free tensor multiplet of its own. Thus, when one
performs DLCQ of this background, one must find KK tower of each
and every one of these separated fivebranes. The  emergence of extra
$L^2$ ground states is thus precisely what we need to attribute them
to free tensor multiplet.\footnote{Another interesting question arises here.
Whether and how these extra $m-1=n+1$ states might be related to
the extra states, $H^{2n+2l}_{compact}=Z$ for $l=2,\cdots,n+2$,
when the fivebranes are all coincident.}

\section{Quantum Screening of Non-Abelian Monopoles}

As mentioned in Introduction, Calabi manifold also makes an appearance in the
context of non-Abelian monopoles \cite{LWY}.
Monopoles in question are those that arise
when $SU(n+4)$ is broken to $U(1)\times SU(n+2)\times U(1)$ and carry either
of the $U(1)$ charges as well as non-Abelian $SU(n+2)$ charges individually.
One could write down family of solution involving one of each, whose combined
asymptotic field has no $SU(n+2)$ charge. The low energy dynamics of such
a soliton pair is described by 4 center of mass, thus free, degrees of
freedom and additional $4(n+2)$ interacting ones. The metric for such
moduli space has been derived in Ref.\ \cite{LWY}, which we present below.

\subsection{Moduli Space of Non-Abelian Monopoles}

Consider $SU(n+4)$ spontaneously broken to $U(1)\times SU(n+2)\times U(1)$.
Of $n+3$
possible fundamental monopoles, all but the first and the last would become
massless, whose degrees of freedom is known to appear in the relative low
energy interaction of the two massive monopoles. Twowo massive monopoles
is charged with respect to each $U(1)$, and both are charged under the
$SU(n+2)$.

The relative moduli space of the monopoles has the topology of $R^{4(n+2)}$.
With the reduced mass of the two massive monopoles,
\begin{equation}
\bar\mu\equiv\frac{m_1 m_{n+3}}{m_1+m_{n+3}}\, ,
\end{equation}
the metric is \cite{LWY}
\begin{equation}
{\cal G}_{\rm rel}=\frac{g^2}{8\pi}
{\cal G}_0+\bar\mu\left(\sum_A d{\vec x}_A \right)^2-
\frac{g^2\bar\mu}{g^2+8\pi\bar\mu\sum_B x_B}
\left(\sum_A x_A\,(d\psi_A+\cos\theta_Ad\phi_A)\right)^2, \label{metric}
\end{equation}
where
\begin{equation}
{\cal G}_0=\sum_A\frac{1}{x_A}\,d{\vec x}_A^2 +x_A
(d\psi_A+\cos\theta_Ad\phi_A)^2
\end{equation}
is a flat $R^{4n+8}$ metric. The summations are over $A=1,\dots,n+2$.
The metric is hyper-K\"ahler, as it must be, and the
three independent K\"ahler forms are
\be
w^{(a)}=\frac{g^2}{8\pi}w^{(a)}_0-\frac{\bar \mu}{2}
\epsilon^{abc}\left(\sum_A dx^b_A\right)\wedge \left(\sum_B dx_B^c\right).
\label{3k}
\ee
The magnetic coupling constant $g$ is related to the electric coupling
constant $e$  by $eg=4\pi$.
This space has $SU(n+2)$ triholomorphic Killing vector fields, which
comes from the unbroken $SU(n+2)$ gauge symmetry.

\subsection{Classical Potential}

When there is more than one adjoint Higgs vev turned on, the low
energy dynamics acquires a new set of terms involving static potential.
It has the general form \cite{BLLY,n=4,n=2'},
\be
{\cal V}=\frac12 (G_\mu G^\mu +\nabla_\mu G_\nu \bar\lambda^\mu\sigma_3
\lambda^\nu),
\ee
where $G$ is a linear combination of triholomorphic Killing vector fields
associated with unbroken $U(1)$'s and $\lambda$'s are certain two-component 
fermionic collective coordinates.
In the above metric, the only such
$U(1)$ Killing vector available for $G$ is
\be
\sum_A\frac{\partial}{\partial\psi^A}.
\ee
The bosonic potential is determined by the size $a$ of the second
Higgs vev, and behaves as
\bea
{\cal C}a^2\left(\sum_A x_A -\frac{8\pi\mu}{g^2+8\pi\bar\mu\sum_B x_B}
\left(\sum_A x_A\right)^2\right)
={\cal C}a^2 \frac{g^2L}{g^2+8\pi\mu L},
\eea
with the constant $\cal C$ to be determined and $L\equiv (\sum_B x_B)$.

When the monopoles are separated at large distance, the nontrivial
part of potential behaves as  $\sim 1/L$. Expanding the potential
for large $L$
\be
{\cal V}_B
={\cal C} a^2\times \frac{g^2}{8\pi\bar \mu}\times \frac{1}{1+g^2/8\pi
\bar\mu L}\simeq \frac{\cal C}{4\pi}\times\left( \frac{g^2 a^2}{2\bar \mu}-
\frac{g^2}{16\pi}\frac{g^2a^2}{\bar\mu^2}\frac{1}{L}+\cdots\right).
\ee
BPS mass formulae of the monopoles fixes the value of constant,
${\cal C}=4\pi$, and we find the following asymptotic form of potential,
\be
{\cal V}_B=\frac{g^2 a^2}{2\bar \mu}-
\frac{g^2}{8\pi}\frac{g^2a^2}{2\bar\mu^2}\frac{1}{L}+\cdots.
\ee
Given a fixed separation of the two massive monopoles,
$\vec R=\sum_A{\vec x}_A$,
the minimum value of $L=\sum_A|{\vec x}_A|$ is $R=|\vec R|$.
Thus, classically, there is an attractive potential between the
two massive monopoles whose asymptotic form goes as
\be
({\cal V}_B)|_{minimum}=\frac{g^2 a^2}{2\bar \mu}
-\frac{g^2}{8\pi}\frac{g^2a^2}{2\bar\mu^2}\frac{1}{R}+\cdots.
\ee
One can think of this ``massless'' monopoles settling down at their
classical vacuum, corresponding to being lined up along the
straight line between the two massive monopoles. Quantum mechanically,
on the other hand, they  tends to spread out simply due to
quantum fluctuations. This  increases the effective value of $L$
and soften the attractive potential. This effect is simplest to
observe when  massive monopoles are held at fixed locations, which
we achieve by taking an infinite mass limit while keeping $ga/\mu$ finite.

\subsection{Infinite Mass Limit}

One important difference in low energy dynamics of non-Abelian monopoles is
that not all degrees of freedom are associated with translations
and internal rotations of the massive cores. Rather, one finds additional
non-Abelian long-range degrees of freedoms, which was dubbed as massless
monopole clouds \cite{LWY}. 
In the above example, dynamics of these massless cloud
emerges when we take hyper-K\"ahler quotient with respect to $U(1)$ generated
by $G$, or equivalently
take a limit where the masses of the two massive monopoles goes to infinite.
The resulting metric is
\begin{equation}
\frac{g^2}{8\pi}\,\left(
C_{AB}d{\vec x}_A\cdot d{\vec x}_B + (C^{-1})_{AB} (d\psi_A+\vec \omega_{AC}
\cdot d\vec x_C)(d\psi_B+\vec \omega_{BD} \cdot d\vec x_D)\right) .
\end{equation}
where now $A$ runs from $1$ to $n+1$ (instead of $n+2$). The periodic
coordinate $\psi_A$ has period, $4\pi$, and the symmetric matrix
$C_{AB}$ has the
form,
\begin{equation}
C_{AB}=\left(\frac{\delta_{AB}}{|\vec x_A|} + \frac{1}{
|\sum_{E=1}^{n+1} \vec x_E - \vec R\;|}\right) .
\end{equation}
The last term is common to all
components. The vector potentials $\vec \omega_{AB}$ are again related
to $C_{AB}$ by
\begin{equation}
\vec \nabla_D \,C_{AB}=\vec \nabla_D \times \vec \omega_{AB} .
\end{equation}
This metric has the same form as in (\ref{calabi}), so the relevant
moduli space here is again Calabi manifold.

Furthermore, the quantity $L$
that appears in the classical bosonic potential is now
$\sum_{A=1}^{n+1} x_A+|\sum_{A=1}^{n+1} \vec x_A - \vec R\,|$,
which was previously identified with the coordinate $\rho^2 R$ in
Section 3.
Dynamics of massless monopoles thus inherit the potential of the whole
monopole dynamics, whose bosonic part is,
\be
{\cal V}_B=\left( \frac{g^2 a^2}{2\bar \mu}-
\frac{g^2}{8\pi}\frac{g^2a^2}{2\bar\mu^2}\frac{1}{R\rho^2}+\cdots\right).
\ee
When this potential is deemed to be small, one can treat it as a perturbation
over the purely kinetic dynamics on the Calabi manifold. In particular, the
quantum effective potential for the pair of massive monopoles at
separation $\vec R$ would be given by
\be
\langle \Psi|{\cal V}|\Psi\rangle/\langle \Psi|\Psi\rangle,
\ee
where $\Psi$ is the normalizable ground state of the massless monopoles in the
absence of the potential.

\subsection{Quantum Corrected  Potential}

We would like to estimate the quantum corrected effective potential
between two massive monopoles discussed earlier. In the limit where
the two massive monopoles are infinitely heavy and separated by a fixed
distance $L$, the Hamiltonian is just that of the cloud and is governed
by Calabi manifold of scale $\sqrt{R}$.
The Hamiltonian is decomposed into two parts,
\be
{\cal H}={\cal H}_0 + {\cal V},
\ee
where ${\cal V}$ is the supersymmetric potential of the original
dynamics appropriately reduced. Both term has an implicit dependence
on the distance $R$.

The harmonic form of the previous section is in effect the
ground state wavefunction with respect to ${\cal H}_0$. The quantum
corrected potential is obtained via standard perturbation theory,
\be
V_{eff}(R)=\langle \Psi |{\cal V}|\Psi\rangle /\langle \Psi |\Psi\rangle .
\ee
This is further simplified by noticing that terms involving fermions
in $\cal V$ always change the fermion number by two. On the other hand,
the fermion number is really the degree of the wavefunction expressed
as differential form, and thus any such operator will have vanishing
expectation on any middle form. Thus, the effective potential is found
by evaluating expectation of purely bosonic part
\be
V_{eff}(R)=\langle \Psi |{\cal V}_B|\Psi\rangle/\langle \Psi |\Psi\rangle,
\ee
For evaluation of this, a useful identity is
\be
*(\Psi \wedge *\Psi )= *(\Psi \wedge \Psi )
=k_n\times \left(n(n+4) F_n(\rho)^2 + G_n(\rho)^2\right),
\ee
for some numbers $k_n$. Then, we have
\be
V_{eff}(R)=\langle \Psi |{\cal V}_B|\Psi\rangle/\langle \Psi |\Psi\rangle
= \frac{g^2 a^2}{2\bar \mu}-
\frac{g^2}{8\pi}\frac{g^2a^2}{2\bar\mu^2}\frac{S(n)}{R}
+\cdots,
\ee
where
\be
S(n)\equiv \left(\int \left(n(n+4) F_n(\rho)^2/\rho^2 +
G_n(\rho)^2/\rho^2\right)
\right)/\left(\int \left(n(n+4) F_n(\rho)^2 + G_n(\rho)^2\right)\right) .
\ee
where the integrals are over the Calabi manifold. For large $n$, this
expression approaches zero as,
\be
S(n)\simeq\frac{5}{3n}.
\ee
Thus we find that quantum effect tends to screen the leading attractive
interaction $\sim 1/R$ down to $\sim 1/(nR)$ for large $n$. This is
reminiscent of the screening effect found by Maldacena \cite{Maldacena} and
by Rey and Yee \cite{RY}, from AdS/CFT picture of Wilson line for
quark-anti-quark
pair. The precise behavior of the latter effect does not match up with
the above, which may be attributed to the fact that we are truncating
most of massless non-Abelian degrees of freedom except those associated
with massless monopoles. It is nevertheless interesting that a strong 
screening effect
appears already at the level of the low energy approximation for
monopoles.

\section{Summary}

We considered a supersymmetric sigma model onto Calabi manifold of arbitrary
dimensions, and found an exact, self-dual, square-normalizable, ground state
of the quantum mechanics. This wavefunction is used to show that a
non-commutative instanton of Super Yang-Mills theory in 4+1 dimensions has
a finite sized quantum ground state, and is interpreted as the first KK mode
of DLCQ description of coincident fivebrane theories. Also the same
wavefunction demonstrates how interaction between non-Abelian monopoles
in partially broken Yang-Mills theories experience screening effect at
long range.

The Calabi manifold should play a further role in understanding (2,0) theories.
Physics we probed here are relevant to the free part of fivebrane
dynamics, and interacting (2,0) theory is not really addressed. In the
spirit of DLCQ approach \cite{DLCQ}, one should be able to discover some 
physics
of (2,0) theories from supersymmetric quantum mechanics on the Calabi manifold.
One interesting and immediate question is, for instance, whether the
topological information such as axial and gravitational anomaly
\cite{anomaly} can be computed in such an approach.

\section*{Acknowledgment}

C.K. was supported by the BK21 program of Ministry of education. 
K.L. is supported in part
by KOSEF 1998 interdisciplinary research grant 98-07-02-07-01-5.
This work concluded when one of us (P.Y.) was visiting Aspen 
Center for Physics, and he wishes to thank the Center and  the 
organizers of the workshop, ``Extreme Strings,'' for hospitality.

\section*{Appendix A}
Here we summarize the exterior algebra for left-invariant one-forms obtained
by Cveti\v{c} et.al.\ \cite{CGLP} which is needed to derive (\ref{exterior}).
Let $L_A^B$ be the left-invariant one-forms of $SU(n+2)$, where $L_A^B$ and
$(L_A^B)^\dagger = L_B^A$, which satisfy the exterior algebra
\be
dL_A^B = i L_A^C \; L_C^B.
\ee
Splitting the $SU(n+2)$ index $A$ as $A=(1,2,a)$, one can then identify
$4n+3$ generators of the coset $SU(n+2)/U(n)$ as a real one-form
\be
\lambda = L_1^1 - L_2^2,
\ee
and complex one-forms
\be
\sigma_a = L_1^a,\quad
\Sigma_a = L_2^a,\quad
\nu = L_1^2,
\ee
and their complex conjugates. For these, the exterior algebra reduces to
\bea
d\sigma_a &=& \frac{i}{2} \lambda \sigma_a + i \nu \Sigma_a + \cdots, \nn
d\Sigma_a &=& -\frac{i}{2} \lambda \sigma_a + i\bar\nu \Sigma_a + \cdots, \nn
d\nu      &=& i\lambda \nu + i \sigma_a \bar\Sigma_a, \nn
d\lambda  &=& 2i \nu \bar\nu + i \sigma_a\bar\sigma_a - i\Sigma_a\bar\Sigma_a,
\eea
where $\cdots$ represents the terms lying outside the coset.
Noting that $s,\ p,\ \sigma_a$ and $\bar\Sigma_a$ are holomorphic one-forms,
it is straightforward to obtain (\ref{exterior}) from the above equations.

\section*{Appendix B}

The identity in Eq.~(\ref{identity}) can be obtained in the following way: 
Add an additional pair of doublets, so that we have
\be
\left(\begin{array}{c}\xi_{k} \\ i\zeta_{k} \end{array}\right), \quad k=1,
\dots,n+1,
\ee
and
\be
\left(\begin{array}{c}i\bar \zeta_{k}\\ \bar \xi_{k} \end{array}\right),
\quad k=1,\dots,n+1.
\ee
Furthermore, we extend $U(n)$ to an $U(n+1)$  acting on them as
fundamental and anti-fundamental, respectively.
Starting with the above, we may build a pair of spin $(n+1)/2$ representations
(under $SU(2)_J$) by contracting the $n+1$ doublets in the fundamental of 
$U(n+1)$ with a completely anti-symmetric tensor $\epsilon_{1234\cdots(n+1)}$,
and similarly for those in the anti-fundamental. 

We then build a singlet by multiplying the two spin 
$(n+1)/2$ representations.
Finally we may convert the expression with two $\epsilon$'s into an expression
with $(n+1)$ inner products under $U(n+1)$, thereby arriving at
the right hand side of Eq.~(\ref{identity}) with $B$ and $D$ 
replaced by the same expressions but now with the sums over $k=1,\dots,n+1$.
Call them $\tilde B$ and $\tilde D$, respectively. 
When we take the hypothetical $(n+1)$-th doublets 
to zero, $\tilde B$ reduces to $B$ and $\tilde D$ reduces to $D$.
On the other hand, this makes each of the two spin $(n+1)/2$ quantities vanish
identically since each has to involve a factor of the $(n+1)$-th doublets. 
The spin singlet built from them should vanish as a result, and thus the 
expression on the right hand side of Eq.~(\ref{identity}) has to vanish by 
itself when the sums are taken over $k=1,\dots,n$.

Obviously more such  identities can be obtained 
for invariant $2n+2l$ forms with $l>1$, but these additional 
identities are irrelevant for our purpose.

\section*{Appendix C}
In this appendix we show that the middle form found in the main part is
self-dual. First we observe that
\be
A (\sum_k \xi_k \bar\xi_k)^l (\sum_k \zeta_k \bar\zeta_k)^m
  (\sum_k \xi_k \bar\zeta_k \sum_k \bar\xi_k \zeta_k)^q, \quad l+m+2q =n,
\ee
is dual to
\be
A (\sum_k \xi_k \bar\xi_k)^m (\sum_k \zeta_k \bar\zeta_k)^l
  (\sum_k \xi_k \bar\zeta_k \sum_k \bar\xi_k \zeta_k)^q,
\ee
since the last factor which containing cross terms of $\xi$'s and $\zeta$'s
cannot be generated by $\xi \bar\xi$ or $\zeta \bar\zeta$ combinations.
This can be verified by explicit calculations using Levi-Civita $\epsilon$
tensors and combinatorics. Then it is easy to see that the invariant
middle form
\be \label{selfdual1}
A D^l B^{n-2l}, \quad l = 0, 1, \ldots, [n/2],
\ee
is self-dual since it is expanded in terms of the above forms in a symmetric
way with respect to $\xi$ and $\zeta$.

As far as hodge duality is concerned, we can treat the 1-forms $s$ and $-ip$
on equal footing with $\xi$'s and $\zeta$'s, which leads us to consider
the quantities
\bea
\tilde B &=& \sum_k(\xi_k \bar\xi_k + \zeta_k\bar\zeta_k) + s\bar s - p \bar p
          = B + A, \nn
\tilde D &=& -(\sum_k \xi_k \bar\zeta_k -isp)
              (\sum_k \zeta_k\bar\xi_k + i\bar s \bar p)
             +(\sum_k \xi_k\bar\xi_k +s\bar s)
              (\sum_k \zeta_k\bar\zeta_k - p \bar p)\nn
         &=& D + A^2 + (AB-C)/2.
\eea
Equation (\ref{selfdual1}) implies that $D^l B^{n-2l}$ is self-dual in
$4n$ dimensions spanned by $\xi$'s and $\zeta$'s. In $4n+4$ dimensions,
it means that $\tilde D^l \tilde B^{n-2l+1}$ is self-dual. Expanding it,
we find
\bea
\tilde D^l \tilde B^{n-2l+1}
  &=& D^l B^{n-2l+1}\nn
  &+& (n-2l+1)A D^l B^{n-2l} + \frac{l}{2}A D^{l-1} B^{n-2l+2}\nn
  &+& -\frac{l}{2} C D^{l-1} B^{n-2l+1}\nn
  &+& \frac{l(l+1)}{2}A^2 D^{l-1} B^{n-2l+1}
    + \frac{l}{2}(n-2l+1)A^2 D^{l-1} B^{n-2l+1}\nn
  &+& \frac{1}{2}(n-2l)(n-2l+1)A^2 D^l B^{n-2l-1}.
\eea
Since the terms in the second line is self-dual, we see that
the term linear in $C$ should be self-dual and the term in the first line
is dual to the terms quadratic in $A$. With these dual relations, it is
easy to check that the middle form $\Psi$ is indeed self-dual with the
coefficients in (\ref{coefficient}).

\end{document}